\documentstyle[twocolumn,aps,floats]{revtex}

\begin{document}

\draft \tolerance = 10000

\setcounter{topnumber}{1}
\renewcommand{\topfraction}{0.9}
\renewcommand{\textfraction}{0.1}
\renewcommand{\floatpagefraction}{0.9}
\newcommand{\br}{{\bf r}}

\twocolumn[\hsize\textwidth\columnwidth\hsize\csname
@twocolumnfalse\endcsname

\title{How Much Energy Have Real Fields Time and Space
in Multifractal  Universe?}
\author{L.Ya.Kobelev\\
 Department of  Physics, Urals State University \\ Lenina Ave., 51,
Ekaterinburg 620083, Russia  \\ E-mail: leonid.kobelev@usu.ru} \maketitle

\begin{abstract}
On the base of multifractal theory of time and space ( see
\cite{kob1}-\cite{kob16}) in this paper shown  presence in every space and
time volumes of real space and time fields a huge supply of energy . In
the multifractal Universe every space volume or time interval possesses by
huge amount of energy($\sim10^{60}cm^{3}$)  and we discuss the problem is
it possible this new for mankind sorts of energy to extract. $$ CONTENTS:
$$ 1. Introduction\\2. What are Energy Densities of Real Space and Time
Fields in Multifractal Universe?\\3. How Much Energy Space and Time
Continually Lose? \\4. Where the Loses Energy  Go To? Why We did not
Discovered it Till Now?\\5. Is It Possible to Extract the Energy from
Fields of Time or Space?\\6. Conclusions
\end{abstract}

\pacs{ 01.30.Tt, 05.45, 64.60.A; 00.89.98.02.90.+p.} \vspace{1cm}

]
\section {Introduction}
The fractal model of space and time \cite{kob1}-\cite{kob16} treats the
time and the space with fractional dimensions as real fields. Universe is
formed only by these fields , i.e. our Universe  is fractional material
time and fractional material space and include not any more. As the time
and the space are material fields  with fractional dimensions and
multifractal structure (multifractal sets) they defined on sets of their
carriers of measure. In each time (or space) point ( "points" are approach
for very small intervals of time or space and "intervals" are multifractal
sets with global dimensions for its sets playing role of local dimensions
for Universe in whole ) the dimensions of time (or space) determine
densities of Lagrangians energy for all physical fields ( or new physical
fields for space) in these points. Time and space are binding by relation
$d{\bf r}^{2}-c^{2}dt^{2}=0$ (this relation is only good approach, more
precise relations see at \cite{kob2}). As real fields time and space own
huge supply of energy ( the question about its energy was considered
partly at \cite{kob16}) and these energies may be evaluated. The purpose
of this paper is more detailed  consideration ( in  the mathematical
formalism of multifractal model of time and space presented in
\cite{kob1}-\cite{kob16}) of problem existence huge supply of energy owned
by each element of time and space. The reason having energy lay in
multifractal nature of time and space, i.e. multifractal nature of our
Universe. Time and space  formed Universe and by means of their
multifractal dimensions construct picture  of all physical fields and got
huge amount of energy when Universe was born.   In the \cite{kob16} where
evaluated these energies for case if current time must be turned back. In
this paper we  consider continually loses of these energies by time and
space, evaluate values of loses energies and there is a discussion: may
humankind be provided with part of these energies.

\section{What are energy densities of  time and space real fields in
multifractal  Universe?}

For answer on these questions it is necessary to construct and investigate
equations describing behavior of very small "intervals" of space and time
( in considered theory each minimal interval space or time ( i.e.
multifractal sets) treats as a "point" \cite{kob1}-\cite{kob16}). It is
well known that behavior of objects with very small sizes describes by
quantum laws. Equations for moving such objects are equations describing
diffusion with imaginary coefficient of diffusion. Then equations for
space point $x_{p}$ and time point $t_{p}$ may be written in the frame of
multifractal theory of time and space as quantum equations with
generalized fractional derivatives (GFD)\cite{kob1} -\cite{kob16}.
\begin{equation}\label{1}
 i\hbar D_{+,t}^{1-\varepsilon_{{\bf r}}}x_{p} =m_{{\bf r}}c^{2}x_{p}
\end{equation}
 and
\begin{equation}\label{2}
 i\hbar D_{+,{\bf r}}^{1-\varepsilon_{t}}t_{p}= m_{t}c^{2}t_{p}
\end{equation}
where $m_{{\bf r}}$ and $m_{t}$ are rest masses time and space volumes. In
(\ref{1})-(\ref{2}) we used generalized fractional derivative
$D_{+,t}^{d_{t}}$ and $D_{+,{\bf r}}^{d_{{\bf r}}}$ defined on
multifractal sets (see \cite{kob1}-\cite{kob16}). Following these works we
consider both time and space as the initial real material fields existing
in the world and generating all other physical fields by means of their
fractal dimensions. Assume that every of them consists of a continuous,
but not differentiable bounded set of small intervals (these intervals
further treated as "points"). Consider the set of small time intervals
$S_{t}$ (their sizes may be evaluated in rude approach as Planck sizes).
Let time be defined on multifractal subsets of such intervals, defined on
certain measure carrier $\mathcal{R}^{N}$. Each interval of these subsets
(or "points") is characterized by the fractional (fractal) dimension (FD)
$d_{t}({\mathbf r}(t),t)$ and for different intervals FD are different. In
this case the classical mathematical calculus or fractional (say, Riemann
- Liouville) calculus \cite{sam} can not be applied to describe a small
changes of a continuous function of physical values $f(t)$, defined on
time subsets $S_{t}$, because the fractional exponent depends on the
coordinates and time. Therefore, we have to introduce integral functionals
(both left-sided and right-sided) which are suitable to describe the
dynamics of functions defined on multifractal sets (see
\cite{kob1}-\cite{kob3}). Actually, these functionals are simple and
natural generalization of the Riemann-Liouville fractional derivatives and
integrals:
\begin{equation} \label{3}
D_{+,t}^{d}f(t)=\left( \frac{d}{dt}\right)^{n}\int_{a}^{t}
\frac{f(t^{\prime})dt^{\prime}}{\Gamma
(n-d(t^{\prime}))(t-t^{\prime})^{d(t^{\prime})-n+1}}
\end{equation}
\begin{equation} \label{4}
D_{-,t}^{d}f(t)=(-1)^{n}\left( \frac{d}{dt}\right)
^{n}\int_{t}^{b}\frac{f(t^{\prime})dt^{\prime}}{\Gamma
(n-d(t^{\prime}))(t^{\prime}-t)^{d(t^{\prime})-n+1}}
\end{equation}
where $\Gamma(x)$ is Euler's gamma function, and $a$ and $b$ are some
constants from $[0,\infty)$. In these definitions, as usually, $n=\{d\}+1$
, where $\{d\}$ is the integer part of $d$ if $d\geq 0$ (i.e. $n-1\le
d<n$) and $n=0$ for $d<0$. If $d=const$, the generalized fractional
derivatives (GFD) (\ref{1})-(\ref{2}) coincide with the Riemann -
Liouville fractional derivatives ($d\geq 0$) or fractional integrals
($d<0$). When $d=n+\varepsilon (t),\, \varepsilon (t)\rightarrow 0$, GFD
can be represented by means of integer derivatives and integrals. For
$n=1$, that is, $d=1+\varepsilon$, $\left| \varepsilon \right| <<1$ it is
possible to obtain:
\begin{eqnarray}\label{5}
D_{+,t}^{1+\varepsilon }f({\mathbf r}(t),t)\approx
\frac{\partial}{\partial t} f({\mathbf r}(t),t)+ \nonumber \\ +
a\frac{\partial}{\partial t}\left[\varepsilon (r(t),t)f({\mathbf
r}(t),t)\right]+ \frac{\varepsilon ({\mathbf r}(t),t) f({\mathbf
r}(t),t)}{t}
\end{eqnarray}
where $a$ is a $constant$ and determined by  choice of the rules of
regularization of integrals (\cite{kob1}-\cite{kob2}, \cite{kob7}) (for
more detailed see \cite{kob7}) and the last addendum in the right hand
side of (\ref{5}) is very small. The selection of the rule of
regularization that gives a real additives for usual derivative in
(\ref{3}) yield $a=0.5$ for $d<1$ \cite{kob1}. The functions under
integral sign in (\ref{3})-(\ref{4}) we consider as the generalized
functions defined on the set of the finite functions \cite{gel}. The
notions of GFD, similar to (\ref{3})-(\ref{4}), can also be defined and
for the space variables ${\mathbf r}$. The definitions of GFD
(\ref{3})-(\ref{4}) needs in connections between fractal dimensions of
time $d_{t}({\mathbf r}(t),t)$ and characteristics of physical fields
(say, potentials $\Phi _{i}({\mathbf r}(t),t),\,i=1,2,..)$ or densities of
Lagrangians $L_{i}$) and it was defined in cited works. Following
\cite{kob1}-\cite{kob15}, we define this connection by the relation
\begin{equation} \label{6}
d_{t}({\mathbf r}(t),t)=1+\sum_{i}\beta_{i}L_{i}(\Phi_{i} ({\mathbf
r}(t),t))
\end{equation}
where $L_{i}$ are densities of energy of physical fields, $\beta_{i}$ are
dimensional constants with physical dimension of $[L_{i}]^{-1}$ (it is
worth to choose $\beta _{i}^{\prime}$ in the form $\beta _{i}^{\prime
}=a^{-1}\beta _{i}$ for the sake of independence from regularization
constant). The definition of time as the system of subsets and definition
of the FD for $d_{t}$ (see (\ref{6})) connects the value of fractional
(fractal) dimension $d_{t}(r(t),t)$  with each time instant $t$. The
latter depends both on time $t$ and coordinates ${\mathbf r}$. If
$d_{t}=1$ (an absence of physical fields) the set of time has topological
dimension equal to unity. The multifractal model of time allows ( as was
be shown \cite{kob5}) to consider the divergence of energy of masses
moving with speed of light in the SR theory as the result of the
requirement of rigorous validity of the laws pointed out in the beginning
of this paper in the presence of physical fields (in the multifractal
theory there are only approximate fulfillment of these laws). We bound
consideration only the case when relation $d_{t}=1-\varepsilon({\mathbf
r}(t),t))$, $|\varepsilon|\ll 1$ are fulfilled.  In that case the GFD (as
was shown) may be represented (as a good approach) by ordinary derivatives
and relation (\ref{1}, (\ref{5})) are valid. So the equations (\ref{1})
-(\ref{2}) reeds ( we used for GFD approach of (\ref{5})(\cite{kob16}))
 \begin{equation}\label{7}
 i\hbar \frac{\partial}{\partial t} x_{p} - m_{t}c^{2}x_{p}+
 + i\hbar\frac{\partial}{\partial t}[\varepsilon x_{p}]+
i\hbar \frac{\varepsilon}{t} x_{p} = 0
\end{equation}
\begin{equation}\label{8}
 ci\hbar \frac{\partial}{\partial {\bf r}} t_{p} - m_{{\bf r}}c^{2}t_{p}+
 + ci\hbar\frac{\partial}{\partial \bf r }[\varepsilon t_{p}]+
ci\hbar \frac{\varepsilon}{{\bf r}} t_{p} = 0
\end{equation}
where $c$ -speed of light.  These equations describe behavior of the
volumes with "point" sizes in time and space (we  remind once more that it
is only  the approach that we use and in reality minimal size of time
intervals and minimal sizes of space intervals in the theory  are bound,
for example, by Planck sizes, thou the last are multifractal sets too) For
free volumes choose solutions for $x_{p}$ and $t_{p}$ as a plane waves
with energy depending of time (we consider further only case of $x_{p}$
and omit the members with masses)
\begin{equation}\label{9}
x_{p}=x_{0}\exp\frac{-iE(t)t}{\hbar}
\end{equation}
and for domain of time-space where by members with $\frac{\partial
\varepsilon}{\partial t}$ may neglect ( i.e. fractional additives almost
constant) receive ($\frac{1}{t}= P(\frac{1}{t})-i\pi \delta(t)$, P-mean
value of integral, $\delta(t)$ -$\delta$ function)
\begin{equation}\label{10}
  E(t) =
  -\frac{i\hbar\varepsilon_{t}}{t}+\frac{\hbar\varepsilon_{t}\pi}{t}
\end{equation}
We do not consider the solutions of equation (\ref{8}). If admit relation
$x=ct$ as was admitted in \cite{kob1}-\cite{kob2} the energy of time
"volume" will be of same order that the energy of space volume. Then
(\ref{10}) (or the energy of time volume from the equation (\ref{8}) gives
if fractal dimensions defined by gravitation field ($\varepsilon \sim
10^{-7}, t\sim10^{17}$)
\begin{equation}\label{11}
         E\sim10^{-51}
\end{equation}
This energy belongs to "point" with coordinate $x_{p}$. In the considered
model multifractal time and space each points is multifractal set with
global dimensions $d_{t}$ or $d_{r}$. Let us characterize the volumes of
these points by Planck sizes: $t_{p}\sim10^{-44}sec$,
$x_{p}\sim10^{-33}cm$. The density of energy in the one $cm^{3}$ is equal
\begin{equation}\label{12}
 E\sim10^{60}ev
\end{equation}
So each cubic centimeter of space has such huge energy. This gigantic
energy is determined by fractional dimensions of time in the domain of
space where we live. The energy determined by fractional dimensions of
space ($\varepsilon(t(r),r)$) may be evaluated if find value of the
fractional additives in this case. What is nature of these energies? It
originates by all physical fields born by fractional dimensions of time (
fractional dimensions gravitational, weak-electro-magnetic, strong
interaction and so on and their vacuum ). We took into account only
gravitational field as example. There are vacuum states of all physical
fields in any point of our Universe, they give constant additives at
fractional dimensions (stress difference of vacuum physical fields from
vacuum (carrier of measure) that born our  Universe). Thus  huge energies
of space and time are constructed by the multifractal structure of our
world and are consequences of multifractal nature Universe.

\section{How much energies time and space continually lose?}

The huge supply of energy are constructed by the fractional nature of time
and space . The relation (\ref{10}) (or analogies relation for time
volume) demonstrated diminishing of fractal structure with increasing time
flow and with space  expending. Both time and space tends to the state
where their fractal structures tends to zero and it will the end of
Universe. How much energy time lose each second? The relation (\ref{10})
allows to evaluate its value. We use (\ref{10}) and write
\begin{equation}\label{13}
  \triangle E\sim-\frac{\hbar\varepsilon\tau}{t_{0}^{2}}
\end{equation}
where $\tau=t-t_{0}$ is a current time. So the losees space energy by one
$cm^{3}sec$  are
\begin{equation}\label{14}
  \triangle E\sim 10^{43}evsec
\end{equation}
This is the huge flow of energy (in our case the flow of gravitation field
energy, because we do not consider flows of other fields). Where this
energy going to?

\section { where  the losees energy  go  to? Why we did not discovered it
till now}

It seems only one the answer at the question put above  exist: the flow of
energy are born by diminishing of fractal structure of space and time goes
directly to the carrier of measure that created our Universe. The fractal
theory of time and space now are in study of construction thou main
principle of it formulated and main prediction are made ( see
\cite{kob1}-\cite{kob16}). Our Universe connected (bind) with vacuum
(measure carrier) and continually returns it the energy that got when "big
bang" happened. So Universe must be filled of "radiation" (gravitational,
electro-magnetic and so on energy flows )  that it continually losing. We
mention once yet that in the form of these energy flows the time and the
space (Universe) directly return the energy (it is the energy they had got
from the carrier of measure (physical vacuum) that born them when "big
bang" had happened) back to the carrier of measure (vacuum). Universe
behaves himself as a big reserver of energy strong connected with vacuum
and directly return vacuum her energy. Stress that carrier of measure in
considered model of fractal world not belongs our Universe ( as forest not
belongs Earth till it died). The Universe (Universe consists of  the real
fields of time and the space) only defined on the carrier of measure as
part of it and may be many such Universes defined on the same carrier of
measure. So time an space constantly return energy  to carrier of measure
in the form of energy flows diminishing the fractal structure and
decreases  gigantic energy of all physical fields. We can not see or sense
these energies as we can not sense for example very low frequency of
electro-magnetic field. For experimental discovers of  these flows (time
flows and space flows)  it seems necessary some a new devices based on a
new theory of vacuum that born our Universe and more deep penetration in
the nature of time and space fields. Physics needs more knowledge about
the nature of carrier of measure, characteristics of interactions between
time and space fields and measure carrier. These problems may be solved in
future.

\section{is it possible to extract the energy from fields of
time or space?}

As was shown in the above paragraph the multifractal theory of time and
space the real fields of time and space are filled by energy. The one
$cm^{3}$ only space field is a container of the  huge amount of  energy
because only lose of energy during one second of time consists $\sim
10^{43}ev$. The energy of rest mass is almost nothing as compared with
this energy. This energy transfers directly in the measure carrier (see
above paragraphs). Is it possible extract it and use for practical
purpose? It is very difficult to answer on this question now because there
are many non researched problems in the theory of real fractal time and
space fields. This problem is one of the main problems of the theory . If
somebody asks the Neanderthal man about  possibility to use for needs of
his tribe the energy of water flow of near river what he will answer? He
could not even understand what he was asked about. Now the method of
transferring of the energy of physical fields flows directly at the energy
fields of carrier of measure unknown for mankind. What new devises must be
invented for discover and using these flows of energy? What new physics
needs for it? Nobody can answer on these question now exactly.
Nevertheless  for any energetical flow must exist some of stopping flow
devices that may take some of flow energies. I think early or later such
devices will be invented. For it the new theory of vacuum is necessary  .
Now such the theory of vacuum is absent. It is necessary find out how
correctly describe interactions between time and space flows, physical
fields plows and measure carrier (vacuum). If it will be done (nobody
knows how much time it needs) the stopping (bar) flows devices will be
invented. We suppose that these problems will be solved in future then the
two problem are decided:\\ a)the  problem of sources of any amount of
energy;\\ b) the problem of governing by time, because the taking energy
from time field flow partly stopped or partly inverse the time flow.\\\\
 In the beginning of the last century had appeared the
known relation $E = mc^{2}$. It was not simple understand its influence on
human life. It relation allowed humankind to develop new technology of
atomic era. May be huge amount of energy condensed in the time and the
space fields allow mankind in future era use it too. How to do it nobody
knows now. May be very intensive beams of electro-magnetic field energy
received by means of beams of particles moving with speed of light ( see
\cite{kob4}) allow to do "the opening" in the time field and to receive
from it energy? Is it is possible to use the energy of long wave"
vibrations" that filled Universe? Optimistic point of view allows to have
hope that in future these new sources of energy when energy may be got
directly from time and space fields flows (in case that the fractal theory
of time and space will find experimental verification) may be used.

\section{conclusions}

1.The multifractal theory of time and space forecast existence of gigantic
supply of energy in the real time and space fields (the order of these
losees is $10^{60}evsec$);\\2. The multifractal theory of time and space
forecast existence of gigantic losees of energy by each volumes of space
and time (the order of these losees is $10^{43}evsec$);\\3. Discussion of
possibilities to use the huge supply of time field and space field
energies by humankind gives more optimistic then pessimistic perspective
for future of mankind.

\end{document}